\documentstyle[preprint,revtex]{aps}

\begin{document}
\font\fortssbx=cmssbx10 scaled \magstep2
\hbox{\fortssbx University of Wisconsin - Madison}
\hfill\vbox{\hbox{\bf MAD/PH/708}\hbox{\bf ISJ-4700}\hbox{July 1991}}\par
\vspace{.3in}
\begin{title}
  LONG-WAVELENGTH OSCILLATIONS AND\\ THE GALLEX SOLAR NEUTRINO SIGNAL
\end{title}
\vspace{.4in}
\author{V.~Barger$^1$,   R.~J.~N.~Phillips$^2$  and K.~Whisnant$^3$}
\vspace{.2in}
\begin{instit}
$^1$Physics Department, University of Wisconsin, Madison, WI 53706, USA\\
$^2$Rutherford Appleton Laboratory, Chilton, Didcot, Oxon, England\\
$^3$Physics Department and Ames Laboratory, Iowa State University,
   Ames, IA 50011, USA
\end{instit}
\thispagestyle{empty}

\begin{abstract}
\nonum\section{ABSTRACT}

The recently reported solar neutrino signal in the $^{71}$Ga GALLEX detector
adds a new dimension to the solar neutrino puzzle, complementing the
previously known signals in $^{37}$Cl and water-Cherenkov detectors. Possible
explanations for this new signal in terms of matter-enhanced neutrino
oscillations (MSW effect) are already awaiting in the literature. We point out
here that long-wavelength vacuum oscillations can furnish an alternative
explanation of all three signals simultaneously; such solutions give neutrino
spectra with distinctive energy dependence and seasonal time dependence.

\end{abstract}

\newpage

   The recent observation of a solar neutrino signal in the $^{71}$Ga detector
of the GALLEX group~\cite{gallex1}  has added an important new constraint
in the solar neutrino puzzle.   Going beyond early upper limits and recent more
positive indications
from the SAGE group~\cite{sage}, that uses a different technique based on
metallic gallium,  GALLEX reports a definite signal of $83 \pm 19 \pm 5$~SNU
to be compared with predictions of about 132~SNU in the standard solar model
(SSM)~\cite{ssm,bp} with conventional neutrino propagation.  This indicates a
suppression ratio
\begin{equation}
R_{\rm Ga} = \mbox{(observed Ga rate)/(SSM Ga rate)} = 0.63 \pm 0.16
\label{R_Ga}
\end{equation}
relative to the latest Bahcall-Pinsonneault calculation~\cite{bp},
that differs from the corresponding suppression ratios
\begin{equation}
     R_{\rm Cl} = 0.26 \pm 0.05 \,, \qquad  R_{\rm Kam\,II} = 0.47 \pm 0.09 \,,
\label{R_Cl}
\end{equation}
in the classic $^{37}$Cl Homestake detector~\cite{hom}  and in the  water
Cherenkov ($\nu$-$e$ scattering) Kamiokande\,II detector~\cite{kam}, that have
higher neutrino energy thresholds. Both experimental and theoretical errors are
included here. The new solar neutrino puzzle is to
explain these three different observations simultaneously.

   A first discussion, presented by the GALLEX group itself~\cite{gallex2} and
amplified by others~\cite{gelb,bludman},
argues that an explanation in terms of non-standard solar models is still
at least conceivable, although not particularly promising.
They also show that explanations in terms of matter-enhanced
neutrino oscillations (the MSW effect~\cite{msw}) are possible, for two
distinct regions in the ($\sin^2 2\theta,\, \delta m^2$) parameter plane.
Indeed, several authors have previously studied MSW fits to the Homestake
and Kamiokande\,II data simultaneously via mixing with
active~\cite{barger1,kuo} or sterile~\cite{barger2} neutrino species;  their
solutions broadly agree, their range of predictions for the gallium experiment
exist in the literature and already indicate where the new GALLEX result can be
accommodated in a MSW scenario~\cite{barger1,kuo,barger2}.

   In the present Letter we point out an alternative explanation in terms of
long-wavelength vacuum neutrino
oscillations~\cite{gribov,barger3,glashow,barger4,acker,krastev}; solutions
of this kind~\cite{barger4,acker,krastev}, previously fitted to the Homestake
and Kamiokande data, predict $^{71}$Ga capture rates quite consistent with the
new GALLEX result above.  With such oscillations, having wavelengths
comparable to the Earth-Sun distance, it is natural for some sections of the
solar neutrino spectrum to be greatly suppressed while others suffer
less suppression.  In the following we shall first
present updated long-wavelength oscillation (LWO) fits to the Homestake plus
Kamiokande data, using the most recent version of the SSM~\cite{bp} and
incorporating the first 220~days preliminary results from the upgraded
Kamiokande\,III detector~\cite{nakamura}, that give
\begin{equation}
R_{\rm Kam\,III} = 0.60  {+0.15\atop -0.13} \ . \label{R_KamIII}
\end{equation}
Superposing these solutions on an iso-SNU plot of the corresponding predictions
of a $^{71}$Ga detector exhibits visually the range of GALLEX predictions that
is allowed for this kind of solution and the neutrino mass and mixing
parameters that are required. Finally, we shall present LWO fits to the
Homestake plus Kamiokande plus GALLEX data simultaneously and discuss their
predictions for future observations.

We have first re-fitted the  LWO hypothesis to the latest suppression ratios
from Homestake and Kamiokande\,III (above), together with the Kamiokande\,II
ratios separated into 14~bins of recoil electron energy $T_e$ (their weighted
mean appears in Eq.~(\ref{R_Cl})\,), in order to input the maximal pre-GALLEX
spectral information. The initial solar neutrino spectrum is taken from the
recent Bahcall-Pinsonneault~\cite{bp} update of the SSM, that includes He
diffusion and other improvements on previous calculations~\cite{ssm}. We assume
two-flavor mixing of the electron-neutrino $\nu_e$, either with an active
neutrino species $\nu_\alpha$ ($\alpha=\mu$ or $\tau$) or with a sterile
neutrino $\nu_X$; these two scenarios are indistinguishable in $^{37}$Cl or
$^{71}$Ga detectors, but give different results in detectors (including
Kamiokande) that are sensitive to neutral-current scattering of $\nu_\alpha$.
Figure~\ref{fit} shows our resulting regions of fit in the $(\sin^2
2\theta,\,\delta m^2)$ plane, where $\theta$ is the usual mixing angle and
$\delta m^2$ is the difference of mass-squared eigenvalues. There are 16 data
points (Cl rate, 14~$T_e$ bins
from Kam\,II, Kam\,III rate) and two free parameters; the best fit was for
$\nu_e$-$\nu_\alpha$ oscillations with
$\delta m^2=6.4\times10^{-11}\rm\,eV^2$ and $\sin^2 2\theta=0.83$,
yielding $\chi^2_{\rm min}=12.5$. The regions of fit have summed
$\chi^2<\chi_{\rm min}^2+6.1$ corresponding to 95\%~CL.
As in previous fits~\cite{barger1,barger4,krastev}, we see that
sterile-neutrino mixing solutions are more restricted but not excluded.
We note that $\nu_e$-$\nu_X$ oscillations with maximal mixing are an essential
feature of a recent custom-designed model~\cite{ma} for the controversial
17\,keV neutrino; in such models LWO are then preferable to MSW solutions as
an explanation for the solar neutrino puzzle.

Figure~\ref{fit} also shows time-averaged iso-SNU contours for the $^{71}$Ga
capture rate. We see that the LWO regions of fit to Homestake and Kamiokande
data fall almost entirely between the 60~SNU and 80~SNU contours, predicting
$^{71}$Ga capture rates compatible both with the GALLEX signal of
$83\pm19\pm5$~SNU~\cite{gallex1} and with the published SAGE upper limit of
79~SNU~\cite{sage} at 90\%~CL (compatibility with SAGE data alone was
previously discussed in Ref.~\cite{krastev}). This figure shows that the LWO
hypothesis accommodates the present gallium data quite naturally. It also
shows how a future more precisely determined $^{71}$Ga rate can fit~in.

Finally  we have fitted the LWO hypothesis to all Homestake plus
Kamiokande plus GALLEX results combined (17 data points with two free
parameters); the best fit parameters are nearly identical to those in the
fit without the GALLEX result, and give $\chi^2_{\rm min}=13.7$.
Figure~\ref{combined} shows the corresponding regions of fit at
95\%~CL in the  $(\sin^2 2\theta,\,\delta m^2)$ plane. These regions
summarize the LWO picture for present data.

The LWO predictions for
future experiments are particularly sensitive to line sources in the
solar neutrino spectrum, such as the 862\,keV $^7$Be line  (that generates
most of the wiggles in the $^{71}$Ga contours in Fig.~\ref{fit}).  Observations
of $\nu$-$e$ scattering at the planned BOREXINO  detector~\cite{borexino}, in
the electron recoil energy band $0.26 < T_e < 0.66$\,MeV,
will be very sensitive to this $^7$Be line contribution. Figure~\ref{combined}
shows contours of the time-averaged suppression factor $R$(Borexino) for this
energy band; a range of possible values $0.3 \alt R \alt 0.9$
is allowed for $\nu_e$-$\nu_\alpha$ active neutrino oscillations, or
a range $0.1 \alt R \alt 0.4$  for $\nu_e$-$\nu_X$ sterile neutrino
oscillations.  These are quite wide bands, that fully overlap the range
0.21--0.65 expected for MSW solutions~\cite{gelb,barger1};  unless the
BOREXINO results lie outside the MSW band, or future data make the bands much
narrower, this time-averaged measurement alone will not discriminate
sharply between MSW and LWO solutions. One may also look at higher $T_e$
values, which contain contributions from pep, $^{13}$N and $^{15}$O neutrinos,
but there the number of events is smaller by an order of magnitude and the
statistical uncertainties correspondingly greater.

   A distinctive feature of LWO scenarios, however, is that they contain
clean and potentially resolvable oscillations in the $\nu_e$ survival
probability $P(\nu_e\to\nu_e)=1-\sin^22\theta\sin^2(\delta m^2 L/4E)$,
where $L$ is the distance from source to detector; this feature is absent in
MSW scenarios with larger $\delta m^2$ values where the corresponding
oscillatory factors are averaged due to the size of the solar source.
An immediate consequence is a time dependence of the contributions from line
sources, due to the seasonal changes in the Earth-Sun
distance~\cite{gribov,barger3,glashow,barger4,acker,krastev}; here we fix $E$
and find $L$ dependence in $P(\nu_e\to\nu_e)$. Eventually, it should be
possible to discriminate between LWO and other explanations on this basis
alone, but at present there is little evidence on this score. The $^{37}$Cl
capture rate has a $^7$Be component and could exhibit some time dependence; it
is intriguing to find that our best fit with LWO to the seasonal $^{37}$Cl data
(using results cited in Ref.~\cite{cribier}) is actually better (lower
$\chi^2$) than a fit to constant $R_{\rm Cl}$ with no time dependence. At
present this is just an interesting hint, not a statistically significant
result. The $^{71}$Ga and BOREXINO signals, however, contain larger components
from the $^7$Be line and could provide better evidence.  Typical LWO solutions
with $\delta m^2$ of order $5 \times 10^{-11}$,  $1 \times 10^{-10}$ and $2.5
\times 10^{-10}\rm\,eV^2$ have differences between maximal and minimal
six-month $^{71}$Ga count rate of
up to 8, 17 and 29~SNU, respectively, due to the variation in the Earth-Sun
distance. Ultimately, the statistical uncertainty in a six-month gallium
measurement may be reduced to 7~SNU, so these variations may be detectable in
$^{71}$Ga for solutions with larger $\delta m^2$. In the BOREXINO experiment
the count rate is much higher; the statistical uncertainty in the monthly
measurement of $R$ may be as low as 0.04. The ranges of differences between
maximal and minimal monthly measurements of $R$ in BOREXINO are 0.02--0.24,
0.09--0.45 and 0.39--0.66, respectively, for LWO solutions in the three
aforementioned $\delta m^2$ regions. Hence, there is a strong likelihood that
the time dependence could be observed in BOREXINO in a LWO scenario.

The Sudbury Neutrino Observatory (SNO) experiment~\cite{SNO} cannot detect
$^7$Be neutrinos and will therefore have little time dependence, but will
be able to test a second distinctive property of LWO, namely an
oscillatory modulation of the $^8$B neutrino spectrum. This property
follows immediately from the expression for $P(\nu_e\to\nu_e)$, that
oscillates versus $E_\nu$ when measured at (approximately) constant $L$.
SNO will obtain a determination of the high-energy $^8$B neutrino
spectrum through its measurement
of charged-current $\nu_e d \to p p e^-$ scattering.  Here the neutrino  energy
will be measured directly, not averaged (as in $^{37}$Cl capture) nor smeared
by the recoil electron distribution (as in $\nu$-$e$ scattering).
$P(\nu_e\to\nu_e)$ is given by the ratio of the observed $^8$B spectrum to
the calculated spectrum; the normalization of the latter may be affected by
the solar model but the shape is not. Figure~\ref{Edep} illustrates the
dependence of $P(\nu_e\to\nu_e)$ on $E_\nu$ for our best-fit LWO solution
(including an average over the varying Earth-Sun distance); we
see that a clear oscillation minimum is predicted in the energy range above
5~MeV, the practical threshold for SNO. This behavior is distinguishable from
that of two MSW solutions, also shown. If this $^8$B spectrum modulation or
the $^7$Be time dependence were detected, they would provide the first
case(s) in which a resolved neutrino oscillation had been seen.

SNO will also detect neutral-current $\nu_\alpha d \to \nu_\alpha p n$
scattering, which will help determine if oscillations are occurring to sterile
neutrinos. For example, in a sterile neutrino oscillation scenario both the
CC and NC ranges would be suppressed (to perhaps $R\approx 0.4$), while for
$\nu_e$-$\nu_\alpha$ oscillations only the CC rate would be suppressed.

We conclude the following:

\begingroup \parindent=0pt
(a) The LWO hypothesis with two-neutrino mixing can comfortably account
for the present  $^{37}$Cl, Kamiokande and $^{71}$Ga data.  There are discrete
regions of fit as shown, for either active or sterile neutrino mixing.

(b) Time-averaged BOREXINO measurements may not cleanly discriminate
between LWO and MSW solutions, since their predictions overlap
considerably.

(c) A very distinctive signature of LWO solutions, however, is the seasonal
time-dependence of the $^7$Be line. There is at present no more than an
intriguing hint in the $^{37}$Cl data, but future BOREXINO measurements would
probably be able to detect this seasonal dependence.

(d) Another distinctive LWO signature is the oscillatory modulation of the
$^8$B spectrum shape, which should be tested at SNO. More precise data of all
kinds should also restrict the options in the future.

(e) Measurements of NC scattering in SNO may possibly discriminate between
active-neutrino and sterile-neutrino mixing options.
\par\endgroup

Up until now we have discussed oscillations between two neutrino species,
but oscillations among three neutrino flavors are another possibility.
Although the maximal three-neutrino mixing case (which predicts
$R_{\rm Kam}=0.43$ and a $^{71}$Ga rate of 44 SNU) is clearly disfavored
by the new Kamiokande\,III and GALLEX data, many other scenarios with
mass-squared difference scales in the $\delta m^2\sim 10^{-10}\rm\,eV^2$ range
can comfortably account for these results~\cite{barger1,barger4,acker}.
The allowed range of BOREXINO predictions is larger, and the three-neutrino
solutions have the characteristic seasonal variations and oscillatory
modulation of two-neutrino LWO.

\newpage
\acknowledgements
One of us (VB) thanks the Aspen Center for Physics for hospitality during
the completion of this work. This research was supported in part by the
U.S.~Department of Energy under contract No.~DE-AC02-76ER00881
and contract No.~W-7405-Eng-82, Office of Energy Research
(KA-01-01), Division of High Energy and Nuclear Physics,
and in
part by the University of Wisconsin Research Committee with funds granted by
the Wisconsin  Alumni Research Foundation.

\figure{\label{fit}
 LWO solutions to Homestake plus Kamiokande data are shown
as shaded regions in the $(\sin^2 2\theta,\,\delta m^2)$ plane, for
(a)~$\nu_e$-$\nu_\alpha$  active neutrino mixing ($\alpha=\mu$ or $\tau$) and
(b)~$\nu_e$-$\nu_X$ sterile neutrino mixing.  Solid curves denote
iso-SNU contours of the predicted $^{71}$Ga capture rate in each case.}

\figure{\label{combined}
LWO solutions to Homestake plus Kamiokande plus GALLEX data are
shown as shaded regions in the $(\sin^2 2\theta,\delta m^2)$ plane, for
(a)~$\nu_e$-$\nu_\alpha$ mixing ($\alpha=\mu$ or $\tau$) and
(b)~$\nu_e$-$\nu_X$ mixing.  Solid  curves are contours of the  suppression
ratio $R$ for the time-averaged $\nu$-$e$ scattering signal in the BOREXINO
detector, in the band   $0.25 < T_e < 0.66$\,keV.}

\figure{\label{Edep}
Electron-neutrino survival probability $P(\nu_e\to\nu_e)$ is shown versus
neutrino energy $E_\nu$ for the best-fit
$\nu_e-\nu_\alpha$ LWO solution ($\delta m^2=6.4\times10^{-11}\rm\,eV^2$ and
$\sin^2 2\theta=0.83$, solid curve) and solutions typifying the two MSW regions
of fit:
$\delta m^2=5.0\times10^{-6}\rm\,eV^2$, $\sin^2 2\theta=0.008$ (dashed
curve) and $\delta m^2=1.0\times10^{-5}\rm\,eV^2$, $\sin^2 2\theta=0.8$
(dotted curve).}

\end{document}